\documentstyle[prl,aps,epsf]{revtex}
\advance\textheight by 0.14in
\advance\topmargin by -0.1in
\begin{document}
\draft

\twocolumn[\hsize\textwidth\columnwidth\hsize\csname@twocolumnfalse\endcsname

\title{Stability of the Smectic Quantum Hall State: A Quantitative Study}
 
\author{Hangmo Yi$^1$, H.A.\ Fertig$^1$, and R.\ C{\^o}t{\'e}$^2$}

\address{$^1$ Department of Physics and Astronomy,
 University of Kentucky, Lexington, Kentucky 40506-0055 \\
$^2$ D{\'e}partement de Physique, Universit{\'e} de Sherbrooke, 
 Sherbrooke, Qu{\'e}bec, Canada J1K-2R1}

\date{\today}

\maketitle

\begin{abstract}
We present an effective elastic theory which 
{\em quantitatively} describes the stripe phase of the 
two-dimensional electron gas in high Landau levels ($N\geq2$).  
The dynamical matrix is obtained with remarkably high precision 
from the density-density correlation function  
in the time-dependent Hartree-Fock approximation.  
A renormalization group analysis shows that at $T=0$, 
as the partial filling factor $\Delta\nu\equiv\nu-\lfloor\nu\rfloor$ moves 
away from $1/2$, the anisotropic conducting state 
may undergo quantum phase transitions: 
stripes may get pinned along their conducting direction 
by disorder, or may lock into one another 
to form a two-dimensional crystal.  
The model predicts  
values of $\Delta\nu$ for each transition.   
The transitions should be reflected in the 
temperature dependence of the dissipative conductivity.  
\end{abstract}
]

%\narrowtext

In a strong magnetic field perpendicular to the plane 
of a two-dimensional (2D) electron or hole system, 
the energy spectrum is characterized by discrete 
Landau levels (LLs) separated by the cyclotron 
energy $\hbar\omega_c=\hbar eB/mc$ and the Zeeman energy $g\mu_BB$.  
Since the degeneracy of each 
LL, ${\mathcal N}_\phi$, is proportional to the 
magnetic field, for sufficiently strong 
magnetic fields, only a small 
number of low-lying LLs are occupied.  In this 
situation, the kinetic energy is practically 
quenched, sometimes leading to interesting 
strongly correlated liquid ground states \cite{dasSarmaBook}.  For example, 
it is now very well established that the ground 
state of a sufficiently clean system is given by the 
fractional quantum Hall fluid \cite{tsui82a} if 
the filling factor $\nu={\mathcal N}/{\mathcal N}_\phi$ 
(${\mathcal N}=\text{number of electrons or holes}$), 
which measures how many LLs are filled, is close to 
some rational numbers such as $1/3$, $1/5$, $2/3$, $2/5$, etc.  

However, recent experiments on high-mobility 
2D electron \cite{lilly99a+b+cooper99a+du99a+pan99a} 
and hole \cite{shayegan99a+papadakis99a} systems have revealed 
that the transport properties are qualitatively different 
for higher filling factors ($\nu>4$ for electrons and $\nu>2$ 
for holes.)  One of the most remarkable findings is that 
the longitudinal resistivities are highly anisotropic near 
half integer filling factors.  As a natural explanation, 
it has been suggested that the electrons form unidirectional 
charge density waves or ``stripes'', predicted earlier 
by Koulakov {\it et al.} \cite{koulakov96a+fogler96a} and 
Moessner and Chalker \cite{moessner96a}.  
An important theoretical development \cite{fradkin99a} 
was the observation that the states of this system 
may be classified by their symmetries, which are highly 
analogous to those of liquid crystals.  The possible 
states include stripe crystals, smectic, and nematic 
phases \cite{smectic}.  Of these, the smectic shows 
promise of explaining the anisotropic transport data 
at very low temperature \cite{nematic}. 
 
While some progress has been made in understanding 
the temperature dependence of transport in this 
system \cite{fradkin99b}, the true groundstate 
when quantum fluctuations are included remains 
a subject of debate.  It has been 
shown\cite{fertig99b,fradkin99a,emery00a} that 
continuous quantum phase transitions may occur 
among different possible states when the system parameters 
are varied.  To assess whether this happens 
requires a knowledge of the parameters 
entering the effective theory from microscopic 
calculations.  In one such study, 
MacDonald and Fisher, using an elastic edge state model, 
found that both interstripe locking interactions 
\cite{fertig99b} and pinning by weak random disorder 
are relevant perturbations at all $\Delta\nu$ \cite{macdonald99a}.  
However, they argued that both perturbations are extremely 
small in any experimentally accessible systems, 
and computed the anisotropic resistivities 
using a semiclassical Boltzmann transport theory.  

\begin{figure}
\epsfxsize=1.8in
\centerline{ \epsffile{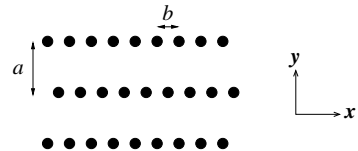} }
\caption{The anisotropic lattice used in the effective 
elastic model.  
}
\label{fig:lattice}
\end{figure}

In this work, we focus on filling factors away 
from $\Delta\nu = 1/2$.  In recent work \cite{cote00a}, 
it was found 
that in the Hartree-Fock approximation, uniform 
stripe states are unstable to the formation of density 
modulations along the stripes {\textemdash} i.e., to 
the formation of a stripe crystal.  
Interestingly, the collective modes \cite{cote00a} 
of the modulated stripe state obtained from a time-dependent 
Hartree-Fock approximation (TDHFA) \cite{cote90a+91a} 
shows that the motion of density deformations 
in the low energy modes \cite{phononModeMovie} looks strikingly 
similar to that of phonon modes in a highly anisotropic lattice 
(see Fig.\ \ref{fig:lattice}).  
Motivated by this observation, we develop an effective 
elastic lattice model which provides a 
{\it quantitative} description 
of the low temperature behavior of the stripe phase.  
As described below, from 
the density-density correlation function in the 
TDHFA, we may numerically obtain 
a dynamical matrix which reproduces the 
low-energy microscopic 
behavior of the system with remarkably high precision.  
Using the result, we  
perform a renormalization group (RG) analysis 
on two important perturbations: interactions 
among stripe modulations, which presumably stabilize 
a stripe crystal, and disorder, which may pin 
the electrons and render the smectic insulating.  
In contrast to uniform stripes \cite{macdonald99a}, 
we find for modulated stripes that, {\it both} 
these perturbations are irrelevant 
close to $\Delta\nu = 1/2$, and that they 
{\it become} relevant at different critical values 
as $\Delta\nu$ moves away from 1/2.  
Either of these Kosterlitz-Thouless (KT) transitions 
represent a metal-insulator transition, and could 
be identified experimentally by measuring 
the filling factor dependence 
of the activation gap on the insulating (i.e., quantized 
Hall effect) side of the transition.  

In our simple lattice model, we assume that each lattice 
site ${\mathbf R}$ is occupied by an object with 
a local density profile or a ``form factor'' 
$f({\mathbf r})$, and that the only dynamic 
variable is the displacement ${\mathbf u}({\mathbf R})$ 
of the object center.  
The elastic energy associated with the displacements 
may be described by a Hamiltonian in Fourier 
space as 
\begin{equation}
H_0 = \frac{1}{2} \sum_{\mathbf q}\sum_{\alpha\beta} u_\alpha^\dagger({\mathbf q}) D_{\alpha\beta}({\mathbf q}) u_\beta({\mathbf q}), \quad (\alpha,\beta=x,y). \label{eq:H0}
\end{equation}
Since the Hamiltonian is Hermitian and invariant under 
a rotation by $\pi$, all dynamical density matrix 
elements are real and $D_{xy}({\mathbf q})=D_{yx}({\mathbf q})$, 
leaving us only three independent real 
parameters per wave vector.  
Below, we will 
treat $D_{\alpha\beta}$ as a fitting parameter, 
chosen to match the density-density 
correlation function \cite{he93a} obtained from the TDHFA of 
the underlying microscopic fermion 
model in the partially filled uppermost LL.  

Since the density-density correlation function is 
computed at the one-phonon level in the TDHFA \cite{cote90a+91a,cote00a}, 
a proper matching procedure requires 
the corresponding quantity 
in the lattice theory to be defined also within 
a harmonic approximation.  More specifically, 
assuming ${\mathbf q}\cdot{\mathbf u}$ is small, 
the density fluctuation operator in the 
lattice model may be written as \cite{giamarchi95a} 
\begin{eqnarray}
& & \delta n({\mathbf q}) \equiv n({\mathbf q}) - \left< n({\mathbf q}) \right> \nonumber \\
& & \quad \approx f({\mathbf q}) \biggl[ - i{\mathbf q}\cdot{\mathbf u}({\mathbf q}) + \sum_{\mathbf G\neq0} \int d{\mathbf r}\,e^{i{\mathbf G}\cdot[{\mathbf r}-{\mathbf u}({\mathbf r})]-i{\mathbf q}\cdot{\mathbf r}} \biggr], \label{eq:density}
\end{eqnarray}
where ${\mathbf G}$ is a reciprocal lattice 
vector and $\langle\cdots\rangle$ denotes an average over the above 
elastic Hamiltonian.  
Ignoring the higher-order second term, 
the density-density correlation function 
may be written as 
\begin{eqnarray}
& & \chi({\mathbf q},{\mathbf q}',\tau) \equiv - \left< {\mathcal T} \delta n({\mathbf q},\tau) \delta n(-{\mathbf q}',0) \right> \nonumber \\
& & \qquad = - \frac{f({\mathbf q}) f^*({\mathbf q}') }{({\mathcal N}ab)^2} \sum_{\alpha\beta} q_\alpha q_\beta' \left< {\mathcal T} u_\alpha({\mathbf q},\tau) u_\beta^\dagger({\mathbf q}',0) \right>, \label{eq:chi}
\end{eqnarray}
where $\mathcal T$ is the time ordering operator and 
$a$ and $b$ are the lattice constants as in Fig.\ \ref{fig:lattice}.  
Due to the lattice symmetry, $u_\alpha({\mathbf q})$ 
is defined only within the first Brillouin zone, 
and the above expression 
vanishes unless ${\mathbf q}'={\mathbf q}+{\mathbf G}$ 
for any reciprocal lattice vector ${\mathbf G}$.  

Using the fact that the single-LL-projected 
displacement operators obey 
$[u_x({\mathbf q}),u_y({\mathbf q}')] = -il_B^2 \delta_{{\mathbf q}+{\mathbf q}',0}$ 
($l_B=\sqrt{\hbar c/eB}$ is the magnetic length), 
the above correlation function 
is easily computed, and may be written in terms 
of the Matsubara frequency $i\omega_n$ as 
\begin{eqnarray}
& & \chi({\mathbf q}+{\mathbf G},{\mathbf q}+{\mathbf G}',i\omega_n) = \nonumber \\
& & \qquad\qquad\qquad \frac{W({\mathbf q};{\mathbf G},{\mathbf G}')}{i\omega_n-\omega_{\mathbf q}} - \frac{W^*({\mathbf q};{\mathbf G},{\mathbf G}')}{i\omega_n+\omega_{\mathbf q}},
\end{eqnarray}
where $W({\mathbf q};{\mathbf G},{\mathbf G}')$ are 
weights that depend on $D_{\alpha\beta}$, 
and $\omega_{\mathbf q} = l_B^2\sqrt{D_{xx}D_{yy}-D_{xy}^2}$ 
is the eigenmode frequency.  

\begin{figure}
\epsfxsize=2.3in
\centerline{ \epsffile{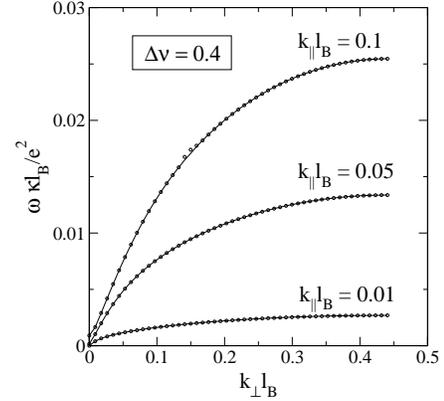} }
\caption{The phonon eigen frequency for $\Delta\nu=0.4$ at several 
wave vectors.  The solid line is the result from the 
TDHFA and the circles are the result of fitting 
using the lattice model.  
}
\label{fig:fitting}
\end{figure}

For each wave vector ${\mathbf q}$ in the first Brillouin 
zone, we need to compute $\chi$ 
for at least three reciprocal lattice vectors 
to fit all unknown parameters in the 
problem.  Assuming the stripes are along 
the $x$-axis as in Fig.\ \ref{fig:lattice}, we use \cite{wavevector} 
${\mathbf G},{\mathbf G}'=0,\pm{\mathbf G}_0$ 
(${\mathbf G}_0\equiv\hat{\mathbf y}2\pi/a$), 
to find the three dynamical 
matrix elements $D_{xx}$, $D_{xy}$, and $D_{yy}$, 
as well as the three form factors $f({\mathbf q})$, 
and $f({\mathbf q}\pm{\mathbf G}_0)$.  
(Due to the inversion symmetry,  
the $f$'s are real.)  
For concreteness, we set $\lfloor\nu\rfloor=6$ (the third lowest LL) 
in our calculations.  In our matching procedure, we 
focus on the $3\times 3$ Hermitian matrix 
$W({\mathbf q};{\mathbf G},{\mathbf G}')$ 
with ${\mathbf G},{\mathbf G}'=0,\pm{\mathbf G}_0$ 
provided to us by the TDHFA.  Together with 
the mode frequency $\omega_{\mathbf q}$, 
this gives ten parameters, of 
which we use six to produce the 
dynamical matrix and form factors.  
(In practice, it is easiest to use the off-diagonal 
elements of the Hermitian matrix.)  
The remaining four 
are used to estimate an errorbar 
and to examine the validity of 
the fitting.  For example, Fig.\ \ref{fig:fitting} 
shows $\omega_{\mathbf q}$ obtained from the TDHFA 
and the lattice theory.  For all the quantities, 
the agreement is quite impressive.  The relative error 
is typically $\sim 10^{-8}$ unless $\omega_{\mathbf q}$ 
is very small, and always remains below $\sim 10^{-3}$.  

The numerical solution of $D_{\alpha\beta}$ agrees remarkably 
well with a harmonic theory of a 2D charged 
smectic system with long-range Coulomb interaction 
such as discussed in Ref.\ \cite{cote00a}.  
Specifically, the functional forms of the dynamical 
matrix elements may be written for small ${\mathbf q}$ as 
\begin{eqnarray}
D_{xx} & \sim & q_x^2 + q_x^2/q, \nonumber \\
D_{xy} & \sim & q_xq_y + q_xq_y/q, \nonumber \\
D_{yy} & \sim & q_y^2 + q_y^2/q + q_x^4, \nonumber \\
\omega_{\mathbf q} & \sim & q_x \sqrt{(q_y^2+q_x^4)/q}, \label{eq:functional} 
\end{eqnarray}
where the terms proportional to $1/q$ are from 
the long-range Coulomb interaction, the $q_x^4$ term 
from bending energy of each stripe, and 
all the rest from effective short-range 
interactions.  

Note that $\omega_{\mathbf q}=0$ at $q_x=0$.  
In fact, the only true gapless mode of the stripe 
{\em crystal} is at ${\mathbf q}=0$.  
However, the gaps are extremely small if $q_x=0$, 
for any $q_y$ \cite{cote00a}.  (In these modes, 
the stripes slide with respect to one another 
\cite{phononModeMovie}.)  
We are thus able to model the results of the 
TDHFA as a charged smectic, despite the 
formal presence of gaps away from ${\mathbf q}=0$.  
Obviously, these nearly gapless modes greatly 
affect the low temperature behavior; 
the RG exponents depend on all $q_y$'s 
up to the first Brillouin zone boundary.  
Thus, a reliable elastic Hamiltonian must 
match the microscopic system at {\it short} 
wavelengths perpendicular to the stripes.  
Along the stripes, however, 
only the long-wavelength 
behavior is important.

\begin{figure}
\epsfxsize=2.3in
\centerline{ \epsffile{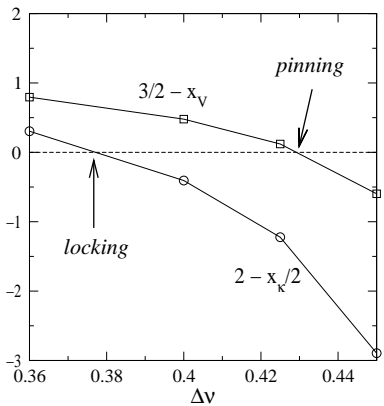} }
\caption{The renormalization group exponents for locking ($2-x_\kappa/2$) 
and pinning along the stripes ($3/2-x_V$).  
The exponents monotonically decreases as $\Delta\nu$ approaches 1/2.  
Errorbars are too small to draw in scale.  
}
\label{fig:exponents}
\end{figure}

Using our quantitative elastic theory, 
we examine some important perturbations and their 
effect at low temperatures by computing the 
RG exponents.  We first 
consider locking between stripes, which controls 
the transition between the smectic and the stripe crystal phases.  
The perturbative action may be written as \cite{fertig99b} 
\begin{eqnarray}
& & S_\lambda = \nonumber \\
& & -\lambda \sum_j \int dx d\tau \cos\frac{2\pi}{b}\left[ u_x(x,j,\tau)-u_x(x,j-1,\tau) \right].
\end{eqnarray}
Here we represent the displacements in real space, 
with the $j$ labeling the stripes.  
For convenience, we have adopted a continuum 
approximation in the $x$-axis, which should not 
affect the small $q_x$ behavior.  A 
standard RG analysis \cite{RG} 
yields the flow equation 
\begin{equation}
\frac{d\lambda}{d\ell} = \left( 2-\frac{x_\kappa}{2} \right) \lambda,
\end{equation}
where 
\begin{equation}
x_\kappa \equiv \frac{2al_B^4}{b} \int dq_y (1-\cos q_ya) \lim_{q_x\to0} \frac{D_{yy}({\mathbf q})q_x}{\omega_{\mathbf q}}. \label{eq:xKappa}
\end{equation}
The exponent is plotted as a function of $\Delta\nu$ 
in Fig.\ \ref{fig:exponents}.  Note that if 
$\Delta\nu\gtrsim0.38$, then $2-x_\kappa/2$ becomes negative 
and locking is irrelevant \cite{instability}.  

Another important perturbation is that 
of weak disorder, 
which may pin the stripes.  We model this with, 
perturbative action for white noise 
disorder, 
\begin{equation}
S_V = \int d{\mathbf r} d\tau\, V({\mathbf r}) n({\mathbf r}),
\end{equation}
where the disorder average is given by 
$\overline{ V({\mathbf r}) V({\mathbf r}') } = V_0^2ab \delta({\mathbf r}-{\mathbf r}')$.  Using the density in Eq.\ (\ref{eq:density}), 
the RG analysis is straightforward.  If the 
${\mathbf G}$ in the second term 
of Eq.\ (\ref{eq:density}) is parallel to the stripes, 
the flow equation for the most relevant 
operator $G_x=2\pi/b$ is given by 
\begin{equation}
\frac{dV_0}{d\ell} = \left(\frac{3}{2}-x_V\right)V_0,
\end{equation}
with 
\begin{equation}
x_V = \frac{al_B^4}{2b} \int dq_y \lim_{q_x\to0} \frac{D_{yy}({\mathbf q})q_x}{\omega_{\mathbf q}}. \label{eq:xV}
\end{equation}
As plotted in Fig.\ \ref{fig:exponents}, 
if $\Delta\nu\gtrsim0.43$, the exponent is negative 
and pinning {\em along the stripes} 
is irrelevant.  

The RG analysis is more complicated 
if ${\mathbf G}$ is perpendicular to the 
stripes.  We find, for any ${\mathbf G}=(0,G_y)$, 
that the free energy 
$F \equiv -\ln \int {\mathcal D}{\mathbf u} \exp (-S_0-S_V)$
computed using Eq.\ (\ref{eq:functional}) 
diverges at low temperatures 
as $T^{-2/5}$ for any $\Delta\nu$.  This 
indicates that pinning {\em across the stripes} is 
always relevant.  Our 
interpretation of this is that the stripes will 
be trapped in channels; however, 
they are still free to move along the 
channels so that this would not spoil 
the phase transitions we found above.  

Using particle-hole symmetry 
for $\Delta\nu>1/2$, the above results imply that 
for $0.43\lesssim\Delta\nu\lesssim0.57$, the stripes are pinned 
only in the the direction perpendicular to 
the stripes at $T=0$, supporting the existence 
of an anisotropic metallic state \cite{fertig99b,emery00a}.  
This is the main result 
of this Letter, and is in qualitative agreement with 
experiment \cite{lilly99a+b+cooper99a+du99a+pan99a}.  

Our results suggest that the anisotropic transport 
properties observed in experiment may well represent 
a new and unusual metallic state of the quantum 
Hall system, separated from insulating (quantized 
Hall) states by quantum phase transitions 
as a function of $\nu$, on either side of $\Delta\nu = 1/2$.  
One direct probe of this is the activation energy $\Delta$ of 
the diagonal transport coefficients $\rho_{xx},\rho_{yy}$ 
in the quantized Hall state.  As the transition is 
approached from below, the gap is controlled by the 
growing correlation length $\xi$, which for a KT transition 
has the characteristic form 
$\Delta \sim \xi^{-1} \sim \exp{\{ -C/\sqrt{|\nu_c-\nu|} \} }$, 
where $C$ is a non-universal constant.  The observation 
of such behavior in finite temperature studies 
{\it away} from $\Delta \nu =1/2$ would constitute direct evidence of 
a new type of phase transition for these systems.  

We conclude with a few final remarks.  
Our simple model was analyzed 
only in the third lowest LL and it 
included neither finite thickness of the 
2D layer nor more complex effects such as 
spin-orbit coupling.  
We expect that the qualitative picture 
presented will be robust, although the 
precise location of the phase transitions 
will surely be affected.  Although in 
our specific model, pinning preempted 
locking, the order of transitions may 
well be reversed when these effects are included.  
We note also that the 
method we have developed here works very well 
away from $\Delta \nu=1/2$, but breaks down 
close to this value.  The breakdown 
occurs for $|\Delta \nu - 1/2| \lesssim 0.01$, and 
is reflected in rapidly growing error bars 
for our matching procedure in that small range.  
This is in part caused by the 
charge motion of the collective modes 
found in TDHFA becoming 
more complicated near half-filling, 
taking on a mixed character of both edge states 
and lattice phonons.  Apparently our simple 
elastic model does not capture this 
complicated behavior.  

In summary, we have developed a method for 
generating a quantitative elastic model of 
quantum Hall stripes from microscopic TDHFA 
calculations.  
An RG analysis shows that an anisotropic conducting 
(smectic) state may be stable against crystallization 
and pinning near filling factor 1/2, and may 
be destabilized by either of these away from 
this filling in a continuous quantum phase transition.  

The authors would like to thank Ganpathy Murthy, 
Allan MacDonald, and Jim Eisenstein for 
helpful suggestions and discussions.  
This work was supported by NSF Grant No. DMR-9870681.


\begin{thebibliography}{10}
\bibitem{smectic}
\label{ref:smectic}
It was argued in Ref.\ \cite{fertig99b} that the stripe crystal can become
  unstable to the proliferation of vacancies and interstitials due to quantum
  fluctuations. If hopping of electrons between stripes is not allowed, the
  resulting state is very likely the same as the smectic.
\bibitem{nematic}
\label{ref:nematic}
Monte Carlo studies of the nematic state in an orientational ordering potential
  (Ref. \cite{fradkin99b}) may explain the temperature dependence of transport
  anisotropy in these systems when $T$ is not too small. However, it is unclear
  whether there is a sharp distinction between the smectic and nematic states
  when such a potential is included.
\bibitem{wavevector}
\label{ref:wavevector}
This choice of wavevector is further justified by the empirical observation
  \cite{phononModeMovie} that, as $q_x\to 0$, the charge motion becomes
  parallel to the stripes and ${\mathbf G} \cdot {\mathbf u}$ remains small.
\bibitem{instability}
\label{ref:instability}
We note that in Ref.\ \cite{cote00a}, the modulated stripes become unstable and
  undergoes a second order phase transition at $\Delta\nu=0.4$, which appears
  to involve the development of stronger modulations along the stripes at lower
  filling factors. While the elastic constants should be continuous through
  this transition, the locking perturbation may increase quickly, so that the
  transition to stripe crystal is likely to be close to $\Delta\nu=0.4$ in our
  model.
\end{thebibliography}
\end{document}